# Программное обеспечение для потоковой обработки фотометрических наблюдений

*В.В. Москвин, А.А. Шляпников*

ФГБУН "Крымская астрофизическая обсерватория РАН", Научный, Крым, 29840
*mvv@craocrimea.ru*

**Аннотация.** Показано использование программных продуктов nova.astrometry.net, SExtractor и Aladin для поиска транзиентных явлений на сериях фотометрических изображений. Рассмотрен алгоритм учета искажений, вносимых в изображения во время наблюдений. Алгоритм основан на корреляционном анализе рядов фотометрических оценок блеска объектов в поле изображения. Показана возможность поиска транзиентных явлений на временах меньших и больших времени наблюдения.

A SOFTWARE FOR STREAMING PROCESSING OF PHOTOMETRIC OBSERVATIONS, *by V.V. Moskvin, A.A. Shlyapnikov*. Software products nova.astrometry.net, SExtractor and Aladin are shown to be used for searching for transient phenomena in series of photometric images. An algorithm for taking into account atmospheric distortions introduced into images during observations is proposed. The algorithm is based on the correlation analysis of the series of photometric estimates of the brightness of objects in the image field. The possibility of searching for transient phenomena is shown to be on time intervals smaller or longer than the observation time.

**Ключевые слова:** анализ данных, обработка изображений, фотометрия

## 1 Введение

Наблюдение красных карликов с активностью солнечного типа – одно из направлений исследований в лаборатории звездного магнетизма Крымской астрофизической обсерватории на протяжении уже полувека (Гершберг, 2017). В настоящее время ведутся наблюдения таких объектов из каталога "GTSh-10" (Гершберг и др., 2011), который содержит 5535 объектов. Они распределены по всей небесной сфере. Есть области с меньшей и большей концентрацией. Примеры областей с большой концентрацией – скопления M42 и M45.

Одновременная регистрация и анализ значительного числа объектов в поле изображения особо актуальны для исследования красных карликов в скоплениях. Наблюдения с панорамным приемником дают возможность получить каталог звездных величин всех объектов в поле кадра. Редукция его в известный фотометрический каталог позволяет проводить анализ интересующих объектов по рядам звездных величин, приведенных в систему фотометрического каталога.

При получении каталога звездных величин необходимо учитывать атмосферные искажения, вносимые в изображения во время наблюдений. Для их учета используются звезды сравнения. Они выбираются методом корреляционного анализа между рядами 10 звезд с минимальной

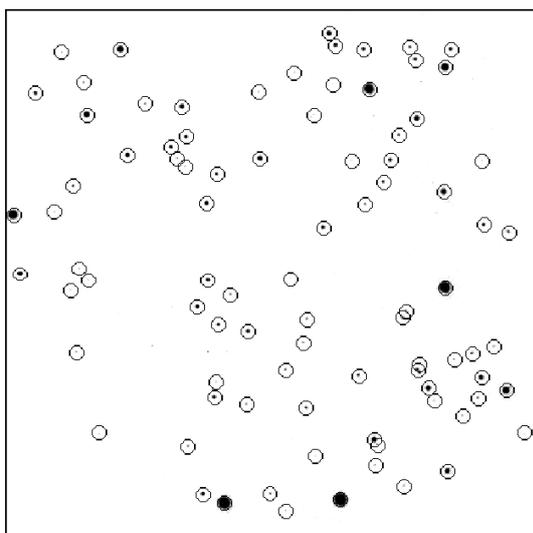

**Рис. 1.** Пример обработки программой SExtractor области переменной V457 Tau с координатами $03^h47^m43^s.7 +24°2'39''$. Угловые размеры изображения $11'×11'$, север – вверху, запад – справа. Пояснение обозначений символов приведены в тексте

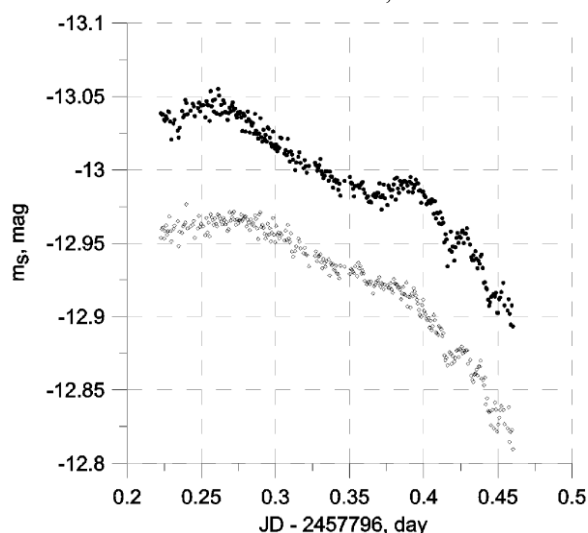

**Рис. 2.** Пример двух кривых блеска без учета атмосферных искажений. $m_s$ – звездная величина, полученная с помощью SExtractor ошибкой. Две звезды с наибольшим коэффициентом парной корреляции выбираются в качестве звезд сравнения. Анализируя динамику изменения блеска этих звезд, находим общие закономерности и учитываем их для всех объектов в кадре. На изображении могут присутствовать звезды, спектральные классы которых не совпадают со спектральными классами звезд сравнения. Поэтому, в общем случае, этот метод можно использовать для поиска транзиентных явлений, но не для точной фотометрической оценки блеска звезды..

## 2 Программное обеспечение

Для решения задачи получения каталога звездных величин из фотометрического астрономического изображения в специализированной литературе найдены такие программные решения, как IRAF (Тоди, 1993), SExtractor (Бертин и Арноутс, 1996), PIXY (Ёсида, 2004). Мы выбрали программу SExtractor, поскольку она используется в стандартных инструментах Международной виртуальной обсерватории (Джорговский, Уильямс, 2005). SExtractor осуществляет поиск объектов точечной или протяженной формы на изображении. На рисунке 1 представлен в качестве примера результат нашего анализа. Светлыми кружочками обозначены объекты, которые обнаружены программой. Точки внутри некоторых кружков обозначают более значительную яркость объектов. В программе реализовано несколько видов апертурной фотометрической оценки блеска.

Для корректного отождествления объектов в поле кадра необходимо произвести астрометрическую калибровку изображения. Для этого используется программа nova.astrometry.net (далее NAN) (Ланг и др., 2010). Данный программный продукт используется удаленно через Интернет, либо устанавливается на компьютер под управлением ОС Linux. Программа сравнивает объекты в исследуемых изображениях с объектами из каталогов USNO-B и 2MASS. Реализована возможность использования других каталогов.

Сравнение каталога объектов, полученного с помощью SExtractor после обработки оригинальных наблюдений, и выбранного фотометрического каталога производится с помощью программы Aladin (Боннарель и др., 2000). Программа является интерактивным атласом неба.

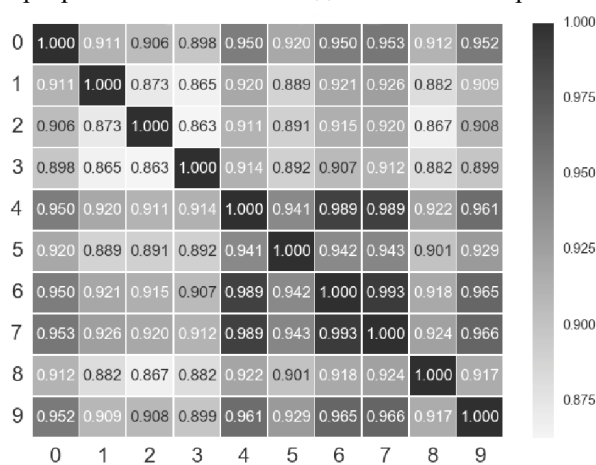

**Рис. 3.** Пример коррелограммы между рядами звездных величин объектов с минимальными ошибками. Строки и столбцы коррелограммы соответствуют 10 звездам с минимальными ошибками, нумерация звезд от 0 до 9. В ячейках таблицы указан коэффициент парной корреляции между рядами двух звезд

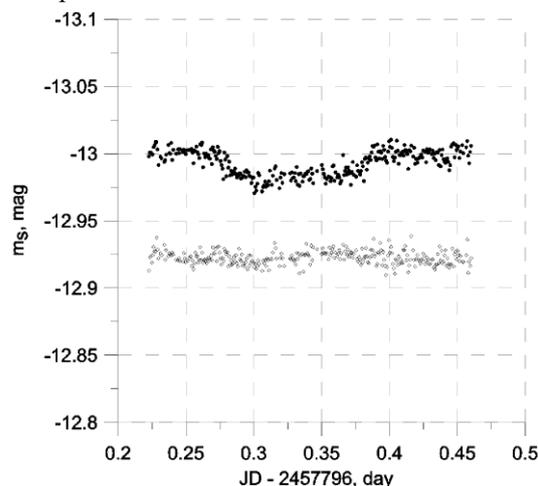

**Рис. 4.** Пример двух кривых блеска после учета атмосферных искажений. $m_s$ – звездная величина, полученная с помощью SExtractor

Она позволяет пользователю визуализировать астрономические изображения, обзоры неба, загружать и работать с различными каталогами небесных объектов.

Программные продукты NAN, SExtractor и Aladin имеют интерфейс работы через командную строку, с помощью которого осуществляется последовательная передача обрабатываемых фотометрических изображений от одной программы к другой. Реализация этой передачи происходит с помощью скрипта, написанного на языке программирования python 3.5. На современных персональных компьютерах обработка 300 кадров формата FITS (Пенс и др., 2010), размером 1024х1024 пикселей, длится около 40 минут.

## 3 Получение каталога звездных величин

Мы проводим апертурную фотометрическую оценку блеска звезд в программном продукте SExtractor. В результате работы связки программ NAN, SExtractor и Aladin, получаем каталоги объектов, которые соотносятся с изображениями. Эти каталоги преобразуются в один сводный каталог, в котором кроме рядов звездных величин приводится их статистика: минимальное, максимальное, среднее значения и среднеквадратическое отклонение.

В течение наблюдений изменяются условия наблюдений, меняется прозрачность и толщина атмосферы в направлении исследуемого объекта. Эти искажения проявляются в виде тренда, характерного для всех объектов в поле кадра (см. рис. 2). Для его учета необходимо выбрать звезды сравнения, проанализировать динамику изменения их блеска и найти общий тренд, а затем учесть его у всех объектов в поле кадра.

Мы используем следующий алгоритм поиска звезд сравнений (Шляпников, 2017). Он основан на том факте, что две звезды не могут синхронно изменять свой блеск. Из набора рядов звездных величин, полученных с помощью SExtractor, мы выбираем 10 звезд с минимальными ошибками. Сравниваем эти ряды на предмет корреляционной зависимости между ними (см. рис. 3). Находим две звезды с наибольшим коэффициентом парной корреляции и берем их как звезды сравнения. По ним определяем общий тренд изменения блеска объектов, после чего вычитаем этот тренд из всех объектов в поле кадра. Рисунки 2 и 4 иллюстрируют изменения, которые происходят с рядами звездных величин при использовании нашего алгоритма. Тренд,

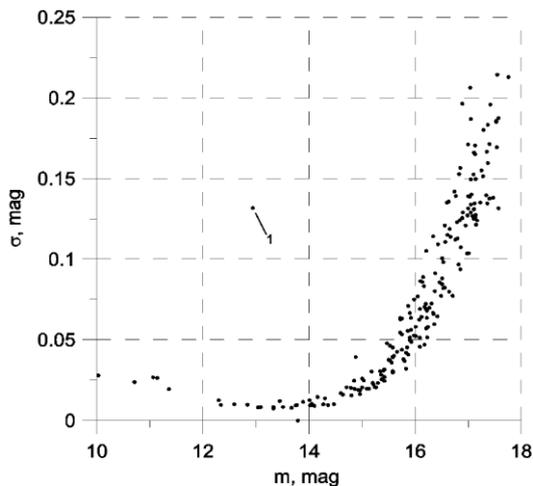 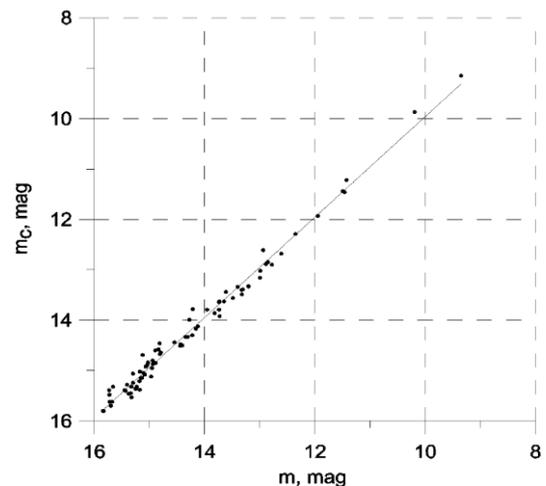

**Рис. 5.** Пример распределения ошибок определения звездных величин (σ). Описание объекта № 1 приведено в тексте

**Рис. 6.** Пример зависимости между звездными величинами, редуцированными в систему выбранного фотометрического каталога (m) и самого каталога ($m_c$) характерный для необработанных кривых блеска, убран, а особенности изменения блеска, характерные для конкретного объекта, остались. На верхней кривой блеска виден провал в центре. Это явление связано с прохождением экзопланеты на фоне звезды.

## 4 Поиск изменений блеска

Объекты в поле изображений могут проявлять изменения блеска на разных временных интервалах. Интервалы могут быть меньше или больше времени наблюдения. Для поиска этих изменений блеска предложены два разных подхода. Изменения блеска со временами меньше экспозиции одного кадра достоверно зафиксировать нельзя.

Для изменений меньших, чем время наблюдения, характерно большое среднеквадратическое отклонение для ряда звездных величин. Оно будет больше, чем среднее значение этого параметра для конкретной звездной величины. Таким образом, для поиска таких явлений можно проанализировать зависимость среднеквадратического отклонения звездных величин для всех объектов от звездной величины (см. рис. 5). Если имеет место изменение блеска у каких-либо объектов в поле кадра, то на графике это будет выглядеть как уклонение одной или нескольких точек из общей зависимости. Эти звезды требует отдельного изучения. В примере на рисунке 5 точка № 1 заметно смещена от общего характера зависимости. Такое поведение объясняется изменением блеска переменной звезды, находящейся в поле кадра (Москвин, 2017).

Для изменений больших, чем время наблюдения, необходимо исследовать звездные величины объектов, полученные с помощью SExtractor и редуцированные в систему выбранного фотометрического каталога. Редукция проводится по звездам с минимальной ошибкой и учетом цветового коэффициента, если наблюдения проводились в нескольких фильтрах, или без его учета, если наблюдения проводились в одном фильтре. После проведения редукции необходимо построить графики зависимости между звездными величинами в редуцированном и фотометрическом каталогах (см. рис. 6). Если звезда имеет блеск, сильно отличающийся от значения по фотометрическому каталогу, то на графике мы будем наблюдать уклонение от общей зависимости и данные объекты будут требовать отдельного изучения. На рисунке 6 приведен пример такой зависимости. В данном случае значительных смещений нет.

# 5 Заключение

Создан скрипт для потоковой обработки фотометрических наблюдений. Он связывает программы NAN, SExtractor и Aladin, получая на выходе каталог астрономических объектов в поле изображения. Скрипт производит учет атмосферных искажений, вносимых в изображения во время наблюдения, а также производит редукцию получившегося каталога в выбранный фотометрический каталог.

По рядам звездных величин, редуцированных в систему каталога, можно осуществлять поиск объектов с транзиентными явлениями и новых переменных. Предложен алгоритм поиска таких явлений длительностью меньше и больше наблюдательного времени.

Скрипт используется для поиска и обработки транзиентных явлений и при обработке наблюдений красных карликов в скоплениях.

## Литература